\documentclass[nofootinbib,aps,pre]{revtex4-2}
\usepackage[utf8]{inputenc}
\usepackage[T1]{fontenc}
\usepackage{amsmath}
\usepackage{amsfonts}
\usepackage{amssymb}
\usepackage{mhchem}
\usepackage{stmaryrd}
\usepackage{xcolor}

\usepackage{graphicx}
\usepackage{tikz}
\usetikzlibrary{arrows.meta}

\newtheorem{theorem}{Theorem}

\newtheorem{condth}{Condition}

\newenvironment{condition}[2][]{\begin{condth}[#1]\label{#2}}{\end{condth}}

\begin{document}

\title{A unified variational framework for conservative and \\nonconservative dynamics of classical systems}

\author{Gabriele Carcassi$^1$, J\'ozsef J. Kolumb\'an$^{2,3}$ and Christine A. Aidala$^1$
}\affiliation{\vbox{1.~Physics Department, University of Michigan, 450 Church Street,
	Ann Arbor, MI 48109-1040,
	United States}\\
  \vbox{2.~Department of Analysis and Operations Research, Institute of Mathematics, Budapest University of Technology and Economics, Műegyetem rkp. 3., H-1111 Budapest, Hungary}
  \vbox{3.~HUN-REN Alfr\'ed R\'enyi Institute of Mathematics, 1053 Budapest, Re\'altanoda utca 13-15, Hungary}
  }

\begin{abstract}
    In previous work we showed how the principle of stationary action can be understood both physically and geometrically as requiring the count of configurations to be conserved over phase space extended by time. In this work we show that the variational principle can be extended to the nonconservative case by noticing that, even in this case, the two-form that quantifies the configuration flux must always give zero contribution along the flow. That is, if $S^a$ is the vector field describing the flow of states, and $w_{ab}$ is the two-form that quantifies the flux of configurations for an independent degree of freedom, then $S^a w_{ab} = 0$ precisely because the flux of configurations must be zero along the flow. This gives a consistent general framework to recast all of classical mechanics, in which we can describe generalized equations of motion in different ways, and understand their connection with the symplectic and contact structures.
\end{abstract}
\maketitle

\section{Introduction}

Given the success of the principle of least action in classical mechanics, and its limitation to conservative systems, many extensions have been proposed and developed to adapt the variational principle to more general contexts. The approaches are rather different. Some supplement the mechanical equations or variational prescription with generalized forces and dissipation functions, as in the Lagrange--d'Alembert and Rayleigh approaches~\cite{gaybalmaz,minguzzi}. Others seek ordinary Lagrangian or Hamiltonian descriptions through explicit time dependence or multiplier methods~\cite{kanai,nuccitamizhmani}, or generalize the derivative dependence of the Lagrangian to include acceleration, higher-order or fractional derivatives~\cite{paisuhlenbeck,riewe,almeidamalinowska}. Another group enlarges the set of variables, introducing auxiliary systems, environmental degrees of freedom or doubled histories~\cite{bateman,caldeiraleggett,galley,galleytsangstein,martindiego}. Herglotz-type approaches instead allow the Lagrangian to depend on the action itself~\cite{lazo,almeidamalinowska}. Geometric extensions include conformally symplectic, contact and cocontact dynamics~\cite{mclachlan,bravetti,cocontact}. Thermodynamic and open-system formulations introduce entropy-production constraints, reversible--irreversible coupling or energy-exchange structures~\cite{gaybalmaz,gaybalmazopen,generic,porthamiltonian}. Finally, surface formulations replace the path-based variational description with one built from a dynamical two-form~\cite{kochanstrings,kochan}.

If we want to understand which one provides a true generalization, we first must answer the question: what does it mean to provide a generalization? As an extreme example, we can say that set theory generalizes the principle of least action, as the latter can be expressed within that larger context given by the former. This is not very helpful as it extends the domain of discourse far beyond classical dynamical systems. Conversely, we can extend the Euler-Lagrange equations using a Rayleigh dissipation function. This is unsatisfactory for the opposite problem: it does not cover all dissipative forces. Therefore, we need to clarify what a system is in the context of classical mechanics, identify a suitable variational principle and then show that it applies to all possible dynamics of such a system, and only such a system. Moreover, the conservative case must be recovered as a special case under a clear additional physical assumption. In the present work we provide such a generalization. The reverse physics approach developed by some of us~\cite{aop-book,aop-rp}, which breaks physical theories into separate mathematical and physical conditions to establish their logical relationships, will supply the methodology of establishing equivalences between different physical and mathematical conditions.

In the context of classical mechanics, the minimum requirement to define a system is to identify a full set of state variables, such as the pairs of position and momentum, and a symplectic form $\omega = \sum_i dq^i \wedge dp_i$ that, at least locally, counts the configurations of each independent degree of freedom (DOF), meaning that it quantifies the range of values for a single pair of conjugate variables in some fixed unit, typically Planck's constant $h$. The counting form, in fact, is needed to define Gibbs entropy for both joint and marginal distributions, without which no invariant notion of reversibility, and therefore of system isolation, can be given. Consequently, it is also tied to the definition of conservative evolution, and therefore of energy.

Given that we are studying the dynamics of the system, the appropriate setting for a generalized variational principle is that of phase space extended by time, as noted in a previous work by some of us that focused on the conservative case~\cite{geoact}. A valid evolution $\gamma(t)$ must satisfy $d_t \gamma^a(t) = S^a$, where $S$ is the displacement vector field that describes the flow of states in the extended phase space. To study the flux of configurations, we define the flux form $w$ which must have two properties: it must recover $\omega$ at equal time, and guarantee that there is no flux along $S$, the direction of the flow (i.e.~$S^a w_{ab} = 0$). This fully characterizes $w$ and guarantees that $S^a$ identifies, up to a scalar factor, the only direction that kills the flux at each point. Therefore,  $$d_t \gamma^a(t) = S^a \iff d_t \gamma^a(t) w_{ab} \eta^b=0$$ for all possible tangent vectors $\eta^a$. This is the key physical and geometric property on which everything else is built. The second condition, in fact, can be expressed in integral form, by requiring that the first variation of the flux between the path and any variation is zero, i.e.~$\delta \phi[\Sigma_{\gamma, \epsilon \eta}]=0$.  In other words, the variational principle is the integral expression of requiring the path to kill the flux form.

In the case of conservative dynamics (DI), the flow preserves the count of configurations in time, i.e.~$d_t \omega =0$, and the flux form is closed and admits, at least locally, a potential one-form $\theta$. Using Stokes' theorem, the condition on the flux can be reformulated in terms of the line integral of $\theta$, which recovers the principle of least action expressed in phase space~\cite{geoact}, i.e.~$\delta \mathcal{A}_H[\gamma] = 0$. Therefore, we have the following relationships:
\begin{equation}
	\begin{aligned}
		d_t \gamma^a(t) = S^a &\iff d_t \gamma^a(t) w_{ab}=0 \iff \delta \phi[\Sigma_{\gamma, \epsilon \eta}]=0 \\&\stackrel{\text{+DI}}{\impliedby} \delta \mathcal{A}_H[\gamma] = 0 \iff d_t \omega =0.
	\end{aligned}
\end{equation}

Under the additional assumption of kinematic equivalence (KE), momentum can be expressed as a function of position, velocity and time, and therefore the variational principle can be fully reformulated in terms of kinematic variables. Restricted by KE, the generalized variational principle can be expressed in terms of a flux functional $\mathcal{F}[x^i_a(t), x^i_b(t)] = \int L_{\mathcal{F}}(x^i_a(t), d_t x^i_a(t), x^i_b(t), d_t x^i_b(t), t)dt$ where $L_{\mathcal{F}}$ is a flux Lagrangian. This recovers a variational principle similar to the one proposed in Refs.~\cite{galley,galleytsangstein}. In the conservative case, we can choose an additive flux Lagrangian, and the principle can be expressed, using Stokes' theorem, on the boundaries. This recovers the standard principle of least action $\delta \mathcal{A}_L[x^i(t)] = 0$ and the action Lagrangian. Summing up, we have:
\begin{equation}
	\begin{aligned}
		d_t \gamma^a(t) &= S^a \stackrel{\text{+KE}}{\impliedby} \delta_b\mathcal{F}[x(t),x(t);\eta(t)]=0 \iff d_t\frac{\partial L_{\mathcal{F}}}{\partial v_b^i}=\frac{\partial L_{\mathcal{F}}}{\partial x_b^i} \\&\stackrel{\text{+DI}}{\impliedby} \delta \mathcal{A}_L[x^i(t)] = 0 \iff L_{\mathcal{F}}^{\mathrm{pr}}(x_a,v_a,x_b,v_b,t)=L(x_b,v_b,t)-L(x_a,v_a, t).
	\end{aligned}
\end{equation}
Note that, since the second path varies around the first, the first variation and derivatives are evaluated on the diagonal.

The key result is that \textbf{all smooth dynamics on phase space can be expressed through the principle of stationary flux and any dynamics expressed through the principle of stationary flux is a dynamics on phase space}. In this sense, assuming smooth dynamics, the principle is exhaustive because it covers exactly all the possible dynamics for classical mechanics. This approach recovers other results, but in a more unified and physically motivated setting. In principle, the treatment can be further generalized by relaxing the smoothness requirement on the dynamics, while retaining the notion of configuration count necessary for classical mechanics. We leave this for future work.

The rest of the paper is structured as follows.  Section~\ref{sec:conditions} provides an overview of the relevant mathematical and physical conditions and the logical relationships among them that will be established.  Section~\ref{sec:dynamicalvar} then establishes the relationships for dynamics described on phase space, and Sec.~\ref{sec:kinematicvar} establishes those for dynamics described in terms of kinematic variables. Section~\ref{sec:othervarprinc} discusses the relationship of our results to other generalizations of the principle of stationary action in classical mechanics, and finally in Sec.~\ref{sec:summary} we summarize the work presented.

\section{Conditions and their relationships}
\label{sec:conditions}

In this section, following the practice of reverse physics, we summarize the mathematical and physical conditions we are going to encounter throughout this work and state their logical relationships within three theorems. The rest of the paper establishes these relationships while explaining their physical meaning and the constructions connecting them.  

\subsection{Background conditions}

Throughout this work, we will assume the following to be true.
\begin{condition}[Smoothness]{SM}
	All geometric objects are assumed smooth.
\end{condition}
Note that, as of now, we do not have an equivalent physical characterization for this condition, and therefore we do not know whether smoothness has physical significance. It is also likely that the theorems established here have more general validity. Therefore this condition, which is often assumed in the context of differential geometry, should be understood as a simplifying assumption.

The following condition characterizes classical states in terms of position and momentum, as is done in the context of Hamiltonian mechanics.
\begin{condition}[Classical states in dynamical variables]{CSD}
	The state space of a classical system with $n$ independent degrees of freedom (DOFs) is represented by position and conjugate momentum. That is, given a set of $n$ position variables $q^i$, we can find a set of $n$ conjugate momentum variables $p_i$ such that each state is identified by position and momentum $(q^i,p_j) \in \mathbb{R}^{2n}$. Moreover, the count of configurations\footnote{By \emph{configuration} we mean the position-momentum pair for a single DOF.} for a DOF is given by $\int_{\Sigma} \omega = \int_{\Sigma} \sum_i dq^i dp_i$ with $\omega = \sum_i dq^i \wedge dp_i$. The count of states is given by the Liouville measure $\int_U \omega^n / n! = \int_U \prod_i dq^i dp_i$. 
\end{condition}
Note that conjugate momentum is not unique in classical mechanics. The relationship between position and conjugate momentum is simply that they, together, allow us to count configurations (i.e.~ways that a single DOF can be specified) through the symplectic form.

The following condition characterizes classical states in terms of position and velocity, as is done in the context of Newtonian and Lagrangian mechanics.
\begin{condition}[Classical states in kinematic variables]{CSK}
	The state \linebreak space of a classical system with $n$ independent degrees of freedom (DOFs) is represented by position and velocity. That is, given a set of $n$ position variables $x^i$, we can find a set of $n$ velocity variables $v^i$ such that each state is identified by position and velocity $(x^i,v^j) \in \mathbb{R}^{2n}$. Moreover, over time evolution, $v^i = d_t x^i$.
\end{condition}
Note that, in this case, no count of configurations is given. However, the condition gives the relationship $v^i = d_t x^i$ between variables during evolution, and no such relationship is given in \ref{CSD}.

The following condition makes sure that we can link the two descriptions.
\begin{condition}[Kinematic Equivalence]{KE}
	The system can be equivalently described in dynamical and kinematic variables. That is, there is an invertible map between position and momentum $(q^i, p_i)$ and position and velocity $(x^i, v^i)$.
\end{condition}
Note that this is an additional condition, as it is independent of \ref{CSD} and \ref{CSK}. However, it is implied by both the principle of stationary action and the principle of stationary flux in their kinematic formulation. That is, the hyperregularity of both the action and flux Lagrangians ensures enough information to recover conjugate momentum and the symplectic configuration count.

\subsection{Evolution and variational conditions}

In the context of \ref{CSD}, i.e.~classical states using dynamical variables, the following three conditions regarding time evolution characterize general dynamics.
\begin{condition}[Equations of motion]{TE-EOM}
	The evolution obeys explicit equations of motion. That is, $d_t \gamma^a(t) = S^a$.
\end{condition}
\begin{condition}[Evolution kills the flux]{TE-FLUX}
	The evolution kills the flux of configurations. That is, there exists a configuration flux form $w_{ab}$ such that $w_{ab}|_{(q,p)}=\omega_{ab}$ and $d_t \gamma^a(t) w_{ab}=0$.
\end{condition}
\begin{condition}[Principle of stationary flux in phase space]{TE-STFL}
	The evolution between two given positions is the one that renders the flux around it stationary. That is, there exists a flux functional $\phi[\Sigma]=\int_{\Sigma} w$ such that $w_{ab}|_{(q,p)}=\omega_{ab}$ and $\delta \phi[\Sigma_{\gamma, \epsilon \eta}]=0$ where $\Sigma_{\gamma, \epsilon \eta}$ is the surface between the path $\gamma^a$ and its variation $\gamma^a + \epsilon \eta^a$.
\end{condition}
We will see that these are actually equivalent characterizations of time evolution, and therefore, following reverse physics standards, they share the same prefix TE. Note that the principle of stationary flux is among the three characterizations.

Still in the context of \ref{CSD}, the following two conditions characterize conservative dynamics.
\begin{condition}[Symplectomorphism]{DI-SYMP}
	The evolution preserves the configuration count and DOF independence. That is, $d_t \omega = 0$.
\end{condition}
\begin{condition}[Principle of stationary action in phase space]{DI-STAC}
	The evolution between two given positions is the one for which the phase space action is stationary. That is, there exists an action functional $\mathcal{A}_H[\gamma] = \int (p_i d_t q^i -H)dt$ such that  $\delta \mathcal{A}_H[\gamma] = 0$.
\end{condition}
In previous work~\cite{geoact}, we have already shown that these two are equivalent, and that, physically, they are equivalent to assuming determinism and reversibility plus the preservation of independence of the degrees of freedom. As such, they share the same prefix DI.

In the context of \ref{CSK}, i.e.~using kinematic variables, we can characterize general dynamics with the following principle.
\begin{condition}[Principle of stationary flux]{KE-STFL}
	The evolution between two given positions is the one that renders the flux around it stationary. That is, there exists a flux functional $$\mathcal{F}[x^i_a(t), x^i_b(t)] = \int L_{\mathcal{F}}(x^i_a(t), d_t x^i_a(t), x^i_b(t), d_t x^i_b(t), t)dt$$ where $L_{\mathcal{F}}$ is a hyperregular flux Lagrangian such that  $\delta_b \mathcal{F}[x^i(t),x^i(t)] = 0$.
\end{condition}
This is effectively condition \ref{TE-STFL} expressed under the assumption of Kinematic Equivalence, which allows us to express momentum, and therefore the flux, in terms of kinematic variables. That is why the prefix is KE.

Lastly, still in the context of \ref{CSK}, we can characterize conservative dynamics with the following.
\begin{condition}[Principle of stationary additive flux]{KEDI-STFL}
	The evolution between two given positions is the one that renders the additive flux around it stationary. That is, there exists a flux functional $$\mathcal{F}[x^i_a(t), x^i_b(t)] = \int L_{\mathcal{F}}(x^i_a(t), d_t x^i_a(t), x^i_b(t), d_t x^i_b(t), t)dt$$ where $L_{\mathcal{F}}$ is an additive hyperregular flux Lagrangian such that  $\delta_b \mathcal{F}[x^i(t),x^i(t)] = 0$.
\end{condition}
\begin{condition}[Principle of stationary action]{KEDI-STAC}
	The evolution between two given positions is the one for which the action is stationary. That is, there exists an action functional $$\mathcal{A}_L[x^i(t)] = \int L(x^i(t), d_t x^i(t), t)dt$$ where $L$ is a hyperregular action Lagrangian such that $\delta \mathcal{A}_L[x^i(t)] = 0$.
\end{condition}
Conservative dynamics is recovered when the flux is additive and the flux Lagrangian can be expressed in terms of the action Lagrangian. Effectively, the principle of stationary action is a special case of the principle of stationary flux for the conservative case. As such, these conditions are labeled KEDI because they capture both KE and DI.

\subsection{Theorems}

The first theorem establishes the correspondence across kinematic and dynamical variables.
\begin{theorem}\label{thrm:bridge}
	Over \ref{SM}, \ref{CSD}+KE+TE-EOM $\iff$ \ref{CSK}+KE-STFL $\impliedby$ \ref{CSD}+KE+DI-SYMP $\iff$ \ref{CSK}+KEDI-STAC.
\end{theorem}
Note how kinematic equivalence (KE) enters differently: for the dynamical variables, it is an extra assumption; for the kinematic variables, it is part of the principles of stationary flux and action. Also note how the conservative case, i.e.~the DI assumption, is a further specialization of the general conditions.

The remaining two theorems establish the characterization for general and conservative dynamics in the respective descriptions.
\begin{theorem}\label{thrm:dyn}
	Over \ref{SM}+\ref{CSD}, TE-EOM $\iff$ TE-FLUX $\iff$ TE-STFL $\impliedby$ DI-SYMP $\iff$ DI-STAC.
\end{theorem}
\begin{theorem}\label{thrm:kin}
	Over \ref{SM}+\ref{CSK}, KE-STFL $\impliedby$ KEDI-STFL $\iff$ KEDI-STAC.
\end{theorem}

Having laid out the various conditions and stated the theorems to be established that capture the logical relationships among these conditions, we now proceed to discuss first dynamics in the extended phase space and then in terms of kinematic variables.

\section{Dynamics in the extended phase space}
\label{sec:dynamicalvar}

In this section we outline the theory for a classical dynamical system on the extended phase space, effectively addressing all equivalences stated in theorem \ref{thrm:dyn}. At each time, we have standard phase space with the geometry given by the count of configurations defined by the closed form $\omega_{ab}$. The dynamics in time, however, need not be conservative (i.e.~Hamiltonian). The central idea is the existence of a flux form $w_{ab}$ that characterizes the flux of configurations through a surface. We find that the only direction that kills said flux form must be the displacement field $S^a$, which defines the flow of states. Intuitively, the flux through surfaces tangent to the flow must be zero. This means that $v^a w_{ab} = 0$ if and only if $v^a = k S^a$ for some $k \in \mathbb{R}$. The principle of stationary flux in general terms is, effectively, an integral way to express the same fact: system evolutions are those paths for which the flux is zero between the path and a small variation. In the conservative case, the dynamics preserves the symplectic form over phase space, the flux form is closed and admits a potential one-form. The action is the line integral of that one-form, and the principle of stationary action, by Stokes' theorem, becomes the principle of stationary flux in phase space formulation.

\subsection{General dynamics}

We consider a system defined by classical mechanics described by dynamical variables, as in condition \ref{CSD}. Each state is fully identified by position and conjugate momentum $(q^i,p_i) \in \mathbb{R}^{2n}$. Given a region $U \subseteq \mathbb{R}^{2n}$, the count of states is given by $\int_U \Omega = \int_U \omega^n / n! = \int_U \prod_i dq^i dp_i$. That is, the volume form $\Omega$ that defines the count of states locally is the product of the count of configurations for each DOF given by $\omega_{ab}$.

Extending this setting with time, states are identified at a specific time using generalized coordinates $\xi^a = (q^i,p_i,t) \in \mathbb{R}^{2n+1}$. A path for the system is defined by a curve $\gamma^a(t) = (q^i(t), p_i(t), t)$, which tells us how the system evolves in time. Note that the time component has to match time itself, $\gamma^t(t) = t$, meaning that time plays a double role of coordinate (i.e.~an element of $\xi^a$) and affine parameter (i.e.~the parameter of the curve). Therefore, a partial derivative $\partial_t$ will correspond to a derivative at $q^i$ and $p_i$ constant, while a total derivative $d_t$ will correspond to a derivative along a path $\gamma^a(t)$.

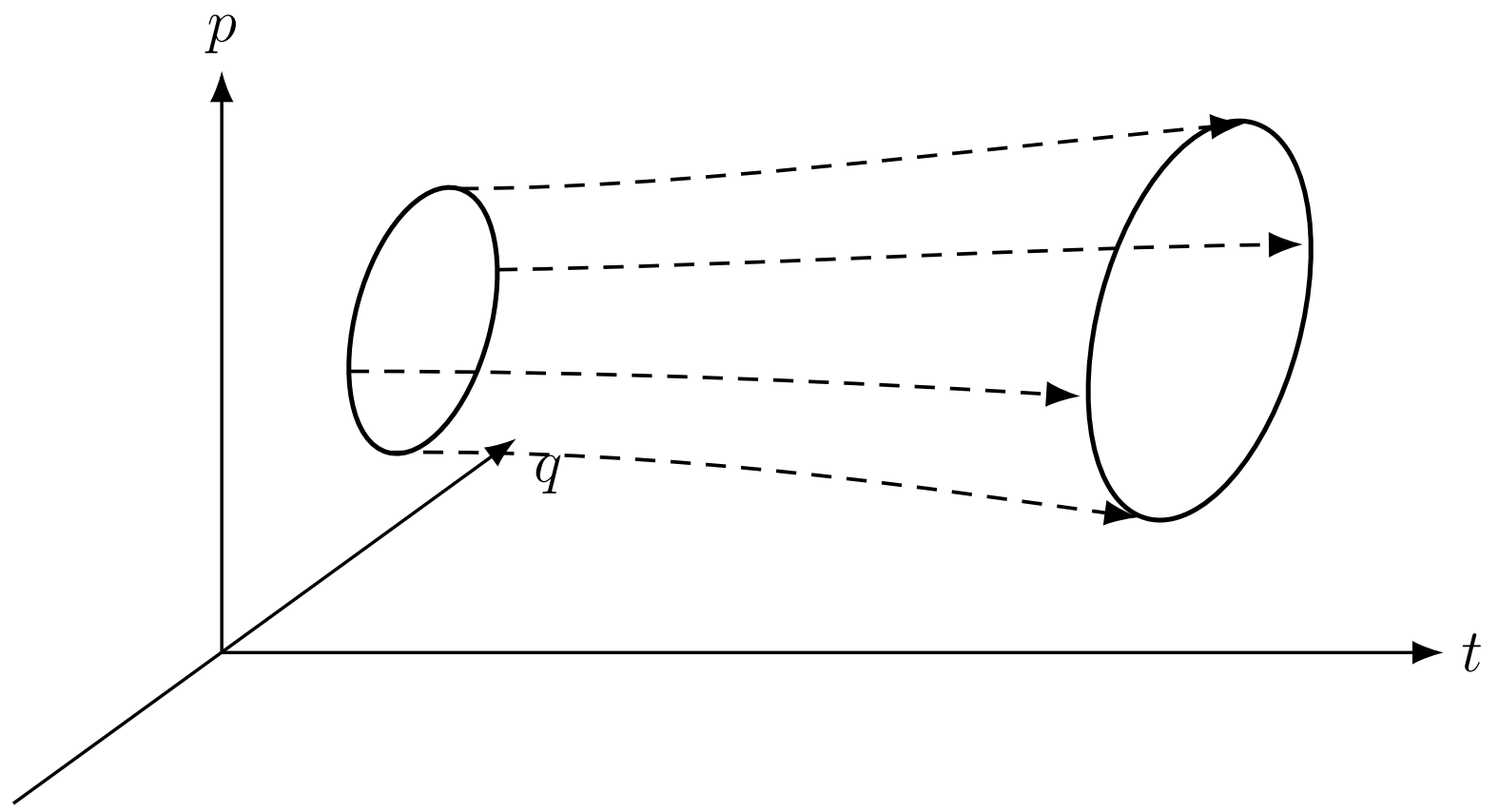
\begin{figure}[h]
    \centering
\begin{tikzpicture}[
    x=1cm,y=1cm,
    axis/.style={line width=0.65pt,-{Latex[length=2.2mm]}},
    section/.style={line width=0.95pt,line join=round},
    flow/.style={line width=0.7pt,dash pattern=on 4pt off 3pt,
                 -{Latex[length=2.4mm]},shorten >=1.5pt},
    every node/.style={font=\large}
]
    \draw[axis] (0,0) -- (8.5,0) node[right] {$t$};
    \draw[axis] (0,0) -- (0,4.05) node[above] {$p$};
    \draw[axis] (-1.45,-1.05) -- (2.05,1.49) node[below right] {$q$};

    \begin{scope}[shift={(1.40,2.31)},rotate=-15]
        \coordinate (aTop)    at (0,0.95);
        \coordinate (aRight)  at ({0.47*cos(30)},{0.95*sin(30)});
        \coordinate (aBottom) at (0,-0.95);
        \coordinate (aLeft)   at ({0.47*cos(210)},{0.95*sin(210)});
    \end{scope}
    \begin{scope}[shift={(6.80,2.31)},rotate=-15]
        \coordinate (bTop)    at (0,1.425);
        \coordinate (bRight)  at ({0.705*cos(30)},{1.425*sin(30)});
        \coordinate (bBottom) at (0,-1.425);
        \coordinate (bLeft)   at ({0.705*cos(210)},{1.425*sin(210)});
    \end{scope}

    \draw[flow] (aTop)    .. controls (3.15,3.23) and (5.05,3.48) .. (bTop);
    \draw[flow] (aLeft)   .. controls (2.65,1.96) and (4.45,1.88) .. (bLeft);
    \draw[flow] (aRight)  .. controls (3.65,2.69) and (5.72,2.82) .. (bRight);
    \draw[flow] (aBottom) .. controls (3.20,1.42) and (5.05,1.12) .. (bBottom);

    \draw[section,shift={(1.40,2.31)},rotate=-15]
        (0,0) ellipse[x radius=0.47,y radius=0.95];
    \draw[section,shift={(6.80,2.31)},rotate=-15]
        (0,0) ellipse[x radius=0.705,y radius=1.425];
\end{tikzpicture}
    \caption{Every two-dimensional surface at constant time is transported by the flow of states to another two-dimensional surface at constant time forming an evolution tube. The flux of configurations must be zero on the sides; therefore, the net flux is the count of configurations at the end minus the count of configurations at the beginning. The dynamics is conservative if the flux over a closed surface is always zero, meaning that each evolution tube conserves the count of configurations and the independence of the degrees of freedom (i.e.~mixed planes, like $(q^1,q^2)$, do not identify configurations for an independent DOF).}\label{evolutionTube}
\end{figure}

Assuming we have a dynamical system, there will be a smooth displacement field $S^a(q^i,p_i,t)$ that defines the flow of states. Its components for each conjugate pair describe the flow of configurations for that DOF. Therefore $\gamma^a(t)$ is an evolution of the system if and only if $d_t \gamma^a(t) = S^a(\gamma^a(t)) = S^a(q^i(t),p_i(t),t)$, which represents the equation of motion. In components, this becomes
\begin{equation}
    d_t \gamma^a(t) = \begin{bmatrix}
\dfrac{dq^i}{dt} \\[1em]
\dfrac{dp_i}{dt} \\[1em]
\dfrac{dt}{dt}
\end{bmatrix} = \begin{bmatrix}
S^{q^i}(q^i,p_i,t) \\[1em]
S^{p_i}(q^i,p_i,t) \\[1em]
S^{t}(q^i,p_i,t) 
\end{bmatrix}= S^a(q^i,p_i,t).
\end{equation}
Since $\gamma^t = t$, we must have that $S^t=1$. This corresponds to condition \ref{TE-EOM}.

We now want to characterize the flux of states through hypersurfaces in extended phase space. As illustrated for configurations in Fig.~\ref{evolutionTube}, if we start with a region $U$ at constant time, this will be mapped by the flow to a region $U'$ at a future time, forming an evolution tube where $U$ and $U'$ are the ends and the sides are swept out by evolution curves. The sides are always tangent to the displacement field $S^a$ and therefore they do not contribute to the flux. Given a generic hypersurface $U$, we are looking for a linear functional $\Phi[U] = \int_U W$, where the state flux form $W$ must coincide with the state counting volume form at equal time, $W|_{(q,p)} = \Omega$, and must be zero along the flow, $W(S^a, \cdot) = 0$. This is enough to characterize the state flux form $W$.

In fact, all components with no time index will either be zero, if there are repetitions, or $\pm 1$ depending on orientation. For example,
\begin{equation}
    \begin{matrix}
        W_{q^1 p_1 q^2 p_2 \cdots q^n p_n} = 1 & W_{q^1 q^2 p_1 p_2 \cdots q^n p_n} = -1 & W_{q^1 q^1 q^2 p_2 \cdots q^n p_n} = 0.
    \end{matrix}
\end{equation}
To find the time components, we can use $W(S^a, \cdots) = 0$, which in component notation is $S^{a^1} W_{a^1 a^2 \cdots a^{2n}} = 0$. If we fix the free indexes with position and momentum variables, the only nonzero terms are those where $a^1$ is either the remaining state variable or $t$, which sets an equivalence between the component of $S^a$ in the remaining state variable and the time component of $W$. For example, suppose $q^i$ is our free variable. The constraint gives us
\begin{equation}
\begin{aligned}
    0 &= S^a W_{\cdots q^{i-1}p_{i-1}\,a\,p_iq^{i+1}p_{i+1}\cdots} \\
    &= S^{q^i} W_{\cdots q^{i-1}p_{i-1}\,q^i\,p_iq^{i+1}p_{i+1}\cdots} + S^t W_{\cdots q^{i-1}p_{i-1}\,t\,p_iq^{i+1}p_{i+1}\cdots} \\
    &= S^{q^i} (1) + (1) W_{\cdots q^{i-1}p_{i-1}\,t\,p_iq^{i+1}p_{i+1}\cdots} \\
\end{aligned}
\end{equation}
\begin{equation}
    W_{\cdots q^{i-1}p_{i-1}\,t\,p_iq^{i+1}p_{i+1}\cdots} = -S^{q^i}
\end{equation}
The time components of the state flux form $W$, then, coincide with the ones of the displacement field up to a sign.

Since $\Omega = \omega^n/n!$, we can also write $W = w^n/n!$ with
\begin{equation}
    \begin{aligned}
    w_{ab} &=
        \begin{bmatrix}
            0 & \delta_i^j & -S^{p_i} \\
            -\delta_j^i & 0 & S^{q^i} \\
            S^{p_j} & -S^{q^j} & 0
        \end{bmatrix} \\
    \end{aligned}.
\end{equation}
The integral $\phi[\Sigma] = \int_{\Sigma} w$ quantifies the flux of configurations through a two-dimensional surface, while $\Phi[U] = \int_U W$ quantifies the flux of states through a $2n$-dimensional hypersurface. Given the setting, then, this is the only natural way to extend the symplectic form to the extended phase space for a dynamical system. This is the configuration flux form described in condition \ref{TE-FLUX}.

Note that the relationship between $S^a$ and $w_{ab}$ is even stronger. Suppose, in fact, that $v^a w_{ab} = 0$. We have
\begin{equation}
    \begin{aligned}
        0 &= v^a w_{ab} = - w_{ba} v^a = - \begin{bmatrix}
            0 & \delta_i^j & -S^{p_i} \\
            -\delta_j^i & 0 & S^{q^i} \\
            S^{p_j} & -S^{q^j} & 0
        \end{bmatrix} \begin{bmatrix}
            v^{q^j} \\ v^{p_j} \\ v^{t} 
        \end{bmatrix} \\
        &= - \begin{bmatrix}
            \delta_i^j v^{p_j} -S^{p_i} v^t  \\ -\delta_j^i v^{q^j} + S^{q^i} v^t \\ S^{p_j}v^{q^j} -S^{q^j}v^{p_j} 
        \end{bmatrix} = - \begin{bmatrix}
            v^{p_i} -S^{p_i} v^t  \\
            - v^{q^i} + S^{q^i} v^t \\
            S^{p_j}v^{q^j} -S^{q^j}v^{p_j} 
        \end{bmatrix} \\
        v^{q^i} &= S^{q^i} v^t \\
        v^{p_i} &= S^{p_i} v^t \\
        S^{p_j}&v^{q^j} -S^{q^j}v^{p_j} = S^{p_j}S^{q^j} v^t -S^{q^j}S^{p_j} v^t = 0,
    \end{aligned}
\end{equation}
where the last equation is not independent of the previous ones and therefore provides no additional constraint. Recalling that $S^t = 1$, we find
\begin{equation}
    v^a w_{ab} = 0 \iff v^a = S^a v^t.
\end{equation}
That is, the displacement field is the only degenerate direction, up to a scalar multiple, of the flux form. This is a consequence of the fact that $w_{ab}|_{(q,p)} = \omega_{ab}$: since $\omega_{ab}$ is nondegenerate, there is only one degenerate direction for $w_{ab}$. This also means that
\begin{equation}\label{evolutionsKillFlux}
    d_t \gamma^a(t) = S^a  \iff d_t \gamma^a(t) w_{ab} = 0.
\end{equation}
That is, $\gamma^a(t)$ is a solution of the equation of motion if and only if it kills the configuration flux form precisely because surfaces tangent to the flow have no flux. This proves the equivalence between \ref{TE-EOM} and \ref{TE-FLUX} over \ref{CSD}. We stress, again, that this relationship is true regardless of whether the dynamics is conservative.

\subsection{The conservative dynamics}

The conservative/Hamiltonian case is exactly recovered when the flux is zero over closed surfaces. Going back to Fig.~\ref{evolutionTube}, if the count of configurations is conserved, flux through every evolution tube is zero. The vanishing flux through infinitesimal evolution tubes, together with the closedness of the equal-time configuration counting form $\omega$, implies that $w$ is closed. We can therefore write $w_{ab} = - \partial_a \wedge \theta_b$.

The spatial parts of $w$ are fixed by $\omega$, and therefore, without loss of generality, we have
\begin{equation}
	\begin{aligned}
		\theta_a &= \begin{bmatrix}
			p_i & 0 & -H
		\end{bmatrix} \\
		w_{ab} &= - \partial_a \wedge \theta_b = \begin{bmatrix}
			0 & \delta_i^j & \partial_{q^i} H \\
			-\delta_j^i & 0 & \partial_{p_i} H \\
			- \partial_{q^j} H & -\partial_{p_j} H & 0
		\end{bmatrix} \\
		S^{q^i} &= \partial_{p_i} H \\
		S^{p_i} &= - \partial_{q^i} H \\
		S^t &= 1.
	\end{aligned}
\end{equation}
This recovers Hamilton's equations. Conversely, Hamilton’s equations give $w=-d\theta$, so $w$ is closed and the configuration count is preserved along every evolution tube. Thus Hamiltonian dynamics is equivalent to condition DI-SYMP.\footnote{In reverse physics, we also find other conditions equivalent to Hamiltonian evolution~\cite{aop-book}.}

The conservative case can equivalently be expressed in terms of the principle of stationary action on the extended phase space. Let us consider a path $\gamma^a(t)$ with $t \in [t_0, t_1]$ and define the action $\mathcal{A}_H[\gamma]$ as the integral of $\theta_a$ along $\gamma^a$. We have
\begin{equation}
    \begin{aligned}
        \mathcal{A}_H[\gamma] &= \int_\gamma \theta_a d\gamma^a = \int_{t_0}^{t_1} \theta_a d_t \gamma^a dt \\
        &= \int_{t_0}^{t_1} \begin{bmatrix}
            p_i & 0 & -H
        \end{bmatrix} \begin{bmatrix}
            d_t q^i(t) \\ d_t p_i(t) \\ d_t t 
        \end{bmatrix} dt
        = \int_{t_0}^{t_1} ( p_i \dot q^i - H) dt.
    \end{aligned}
\end{equation}

Now, let us consider a different path $\tilde \gamma^a(t) = \gamma^a(t) + \epsilon \eta^a(t)$ that starts and ends at the same position. Then $\eta^{q^i}(t_0)=\eta^{q^i}(t_1)=0$, and, since $\gamma^t(t) = t = \tilde\gamma^t(t)$, $\eta^t(t) = 0$. We can take the two paths and close them in a loop with a straight line in momentum at $t_0$ and $t_1$. We can find a surface $\Sigma$ such that its boundary $\partial \Sigma$ is that loop. We can, for example, choose $\Sigma^a_{\gamma, \epsilon \eta} : [t_0, t_1] \times [0,\epsilon] \to \mathbb{R}^{2n+1}$ given by $\Sigma^a_{\gamma, \epsilon \eta}(t,s) = \gamma^a(t) + s \eta^a(t)$, which satisfies $\Sigma^a_{\gamma, \epsilon \eta}(t,0) = \gamma^a(t)$ and $\Sigma^a_{\gamma, \epsilon \eta}(t,\epsilon) = \tilde \gamma^a(t)$. Note that, for a curve $\lambda^a(s) = (\hat{q}^i, p_i(s), \hat{t})$ that only varies in momenta at a fixed position $\hat{q}^i$ and time $\hat{t}$, we have
\begin{equation}
    \int_{\lambda} \theta_a d\lambda^a = \int_{\lambda} \theta_{p_i} d\lambda^{p_i} = \int_{\lambda} 0 d\lambda^{p_i} =0.
\end{equation}
Therefore, the straight lines in momentum provide no contribution to the action. Using Stokes' theorem, we have:
\begin{equation}
    \begin{aligned}
        \mathcal{A}_H[\tilde \gamma] - \mathcal{A}_H[\gamma] &= \oint_{\partial \Sigma_{\gamma, \epsilon \eta}} \theta_a d \gamma^a
        =\iint_{\Sigma_{\gamma, \epsilon \eta}} \partial_a \wedge \theta_b d \Sigma^{ab} \\
        &= - \iint_{\Sigma_{\gamma, \epsilon \eta}} w_{ab} d \Sigma^{ab} = -\phi[\Sigma_{\gamma, \epsilon \eta}]
    \end{aligned}
\end{equation}
That is, the difference of the action between the two paths is minus the flux through the surface between them.

The principle of stationary action, then, amounts to requiring that the flux through the infinitesimal surface connecting a path and any variation is zero. Geometrically, this can only happen if the path is tangent to the displacement field $S^a$, which means it is a solution of the equations of motion. Mathematically, we find:
\begin{equation}\label{variationOnFlux}
	\begin{aligned}
		- \delta \mathcal{A}_H[\gamma] &=
		\delta\phi[\Sigma_{\gamma,\epsilon\eta}]= \partial_\epsilon \phi[\Sigma_{\gamma,\epsilon\eta}]|_{\epsilon=0} \\
		&= \partial_{\epsilon} \int_{t_0}^{t_1} \int_0^{\epsilon} w_{ab} \partial_s \Sigma^a_{\gamma, \epsilon \eta}(t,s) \partial_t \Sigma^b_{\gamma, \epsilon \eta}(t,s) ds dt |_{\epsilon=0} \\
		&= \int_{t_0}^{t_1} \partial_{\epsilon} \int_0^{\epsilon} w_{ab} \eta^a(t) (d_t \gamma^b(t) + s d_t \eta^b(t) ) ds dt |_{\epsilon=0} \\
		&= \int_{t_0}^{t_1} w_{ab} \eta^a(t) (d_t \gamma^b(t) + \epsilon d_t \eta^b(t)) dt |_{\epsilon=0} \\
		&= \int_{t_0}^{t_1} w_{ab} \eta^a(t) d_t \gamma^b(t) dt \\
	\end{aligned}
\end{equation}
Since $w_{ab} \eta^a(t) d_t \gamma^b(t) = 0$ for all position-momentum variations $\eta^a(t)$, $w_{ab} d_t \gamma^b(t) = 0$ whenever $a \in \{q^i, p_i\}$. Now consider $d_t \gamma^a(t) w_{ab} d_t \gamma^b(t)$. This is equal to zero by antisymmetry of $w_{ab}$. Putting the two results together, we have $d_t \gamma^t(t) w_{tb} d_t \gamma^b(t) = 0$. Since $d_t \gamma^t(t) = 1$, $w_{tb} d_t \gamma^b(t) = 0$. Therefore, in the conservative case, 
\begin{equation}
	\delta \mathcal{A}_H[\gamma] = 0 \iff d_t \gamma^a(t) w_{ab} = 0.
\end{equation}
This proves the equivalence between \ref{DI-SYMP} and \ref{DI-STAC}.

\subsection{Stationary flux in the extended phase space}

If we look closely at the derivation in Eq.~\eqref{variationOnFlux}, we note that the only part that depends on conservative dynamics is the expression of the action. The flux and its variation only require $w_{ab}$, which is well-defined for general dynamics. Therefore the following conclusion
\begin{equation}
	\delta \phi[\Sigma_{\gamma, \epsilon \eta}]=0 \iff d_t \gamma^a(t) w_{ab} = 0
\end{equation}
is valid in general. This gives us a way to characterize any dynamics through a variational principle, which can be stated in the following way:
\begin{quote}
	\textbf{Principle of stationary flux.} The path taken by a classical dynamical system between two positions is the one that renders the flux around it stationary.
\end{quote}
This, again, is simply a consequence of the fact that each evolution is locally tangent to the displacement field at every point, and therefore an infinitesimal surface around an evolution will also be tangent to the field, leading to no flux. Conversely, if the path is not tangent, there will be a direction in which a variation will ``catch'' some of the flux. The action exists only in the special conservative case, when the flux form is closed.

To sum up, we have the following five conditions with the following relationship:
\begin{equation}
\begin{aligned}
    d_t \gamma^a(t) = S^a &\iff d_t \gamma^a(t) w_{ab}=0 \iff \delta \phi[\Sigma_{\gamma, \epsilon \eta}]=0 \\&\stackrel{\text{+DI}}{\impliedby} \delta \mathcal{A}_H[\gamma] = 0 \iff d_t \omega =0,
\end{aligned}
\end{equation}
which corresponds to theorem \ref{thrm:dyn}. Once we assume \ref{CSD}, classical states represented by position and momentum, the first three conditions equivalently represent general dynamics while the last two equivalently represent the special case of conservative dynamics.

\section{Dynamics over kinematic variables}
\label{sec:kinematicvar}

In this section we express the principle of stationary flux in kinematic variables, giving a physically motivated variational principle that applies to any dynamics over kinematic variables, effectively addressing all equivalences stated in theorem \ref{thrm:kin}. We define the flux Lagrangian $L_{\mathcal{F}}(x_a, v_a, x_b, v_b, t)$ which is a function of two kinematic states and plays the role of the action Lagrangian for general dynamics. The flux $\mathcal{F}[x^i_a(t), x^i_b(t)]$ is a functional of the two paths that define the contour of the surface. The equation of motion, the conjugate momentum and the symplectic form can all be recovered from the flux and the flux Lagrangian. In the conservative case, the flux Lagrangian, in its proper form, is additive and can be reduced to the difference of an action Lagrangian evaluated on the two states: $L_{\mathcal{F}}(x_a, v_a, x_b, v_b, t) = L(x_b, v_b, t) - L(x_a, v_a, t)$.

\subsection{From dynamical to kinematic picture}

The first task is to re-express the principle of stationary flux in kinematic variables. Additionally, we would like the new principle to work, as much as possible, like the principle of stationary action.

We consider a system as defined by classical mechanics described by kinematic variables, as in condition \ref{CSK}. Each state is fully identified by position and velocity $(x^i, v^i) \in \mathbb{R}^{2n}$, so that, over evolutions, $v^i (t) = d_t x^i(t)$.\footnote{We use $x^i$ and $v^i$, instead of $q^i$ and $\dot{q}^i$, for two reasons. First, to make it clear that these are independent variables used to chart the same state space, and that $v^i (t) = d_t x^i(t)$ applies only on trajectories. Second, to remind ourselves that, while $x^i = q^i$, in general, $\partial_{x^i} \neq \partial_{q^i}$.}

Assuming kinematic equivalence (\ref{KE}), we can change coordinates from $(x^i, v^i)$ to $(q^i, p_i)$ and vice versa. That is, the change of variables
\begin{equation}
	x^i=q^i,
	\qquad
	v^i=S^{q^i}(q,p,t)
\end{equation}
can be inverted to give
\begin{equation}
	q^i=x^i,
	\qquad
	p_i=P_i(x,v,t).
\end{equation}
That is,
\begin{equation}
	S^{q^i}(x,P(x,v,t),t)=v^i.
\end{equation}
For convenience, we also define the force functions
\begin{equation}
	F_i(x,v,t):=S^{p_i}(x,P(x,v,t),t).
\end{equation}
The evolution can therefore be written as
\begin{equation}\label{eq:kinematicEvolution}
	d_t x^i=v^i,
	\qquad
	d_tP_i(x,v,t)=F_i(x,v,t).
\end{equation}

Now, let $x_a(t)$ and $x_b(t)$ be two paths defined on the same interval $[t_0,t_1]$, with velocities $v_a=d_t x_a$ and $v_b=d_t x_b$. Define the differences
\begin{equation}
	\Delta x^i=x_b^i-x_a^i,
	\qquad
	\Delta v^i=v_b^i-v_a^i
\end{equation}
so that we can interpolate between the two paths at each time by
\begin{equation}
	x_s^i=x_a^i+s\Delta x^i,
	\qquad
	v_s^i=v_a^i+s\Delta v^i,
	\qquad s\in[0,1].
\end{equation}
Since $d_t x_s^i=v_s^i$, each intermediate path is a kinematic path. This gives us the following surface on the extended phase space
\begin{equation}
	\Sigma_{ab}(t,s)
	=
	\bigl(x_s(t),P(x_s(t),v_s(t),t),t\bigr).
\end{equation}
This surface corresponds to the two paths in extended phase space connected at every $t$ by a straight segment in kinematic variables.

For brevity, let $P_{i,s}=P_i(x_s,v_s,t)$ and $F_{i,s}=F_i(x_s,v_s,t)$ be the components of the momentum and force functions for $s$. The tangent vectors to this surface are
\begin{equation}
	\partial_s\Sigma_{ab}
	=
	\bigl(\Delta x^i,\partial_sP_{i,s},0\bigr),
	\qquad
	\partial_t\Sigma_{ab}
	=
	\bigl(v_s^i,d_tP_{i,s},1\bigr).
\end{equation}

Applying the configuration flux form to each infinitesimal element, we find
\begin{equation}
	\begin{aligned}
		w_{cd}\partial_s\Sigma_{ab}^{c}\partial_t\Sigma_{ab}^{d}
		&=
		w_{q^i p_j}
		\partial_s\Sigma_{ab}^{q^i}\partial_t\Sigma_{ab}^{p_j}
		+
		w_{q^i t}
		\partial_s\Sigma_{ab}^{q^i}\partial_t\Sigma_{ab}^{t} +
		w_{p_i q^j}
		\partial_s\Sigma_{ab}^{p_i}\partial_t\Sigma_{ab}^{q^j}
		\\
		&\quad+
		w_{p_i t}
		\partial_s\Sigma_{ab}^{p_i}\partial_t\Sigma_{ab}^{t} +
		w_{t q^j}
		\partial_s\Sigma_{ab}^{t}\partial_t\Sigma_{ab}^{q^j}
		+
		w_{t p_j}
		\partial_s\Sigma_{ab}^{t}\partial_t\Sigma_{ab}^{p_j}\\
		&=
		\delta_i^j\,\Delta x^i\,d_tP_{j,s}
		-F_{i,s}\,\Delta x^i\cdot1 -
		\delta_j^i\,\partial_sP_{i,s}\,v_s^j \\
		&\quad+v_s^i\,\partial_sP_{i,s}\cdot1 +
		F_{j,s}\cdot0\cdot v_s^j
		-v_s^j\cdot0\cdot d_tP_{j,s}\\
		&=
		\Delta x^i d_tP_{i,s}
		-F_{i,s}\Delta x^i
		-\partial_sP_{i,s}v_s^i
		+v_s^i\partial_sP_{i,s}\\
		&=
		\Delta x^i\bigl(d_tP_{i,s}-F_{i,s}\bigr).
	\end{aligned}
\end{equation}

We can now express the flux over the whole surface and integrate the first term by parts to simplify the expression.
\begin{equation}
	\begin{aligned}
		\phi[\Sigma_{ab}]
		&= \int_{\Sigma_{ab}} w = 
		\int_{t_0}^{t_1}\int_0^1
		\Delta x^i\bigl(d_tP_{i,s}-F_{i,s}\bigr)\,ds\,dt\\
		&=
		\left[
		\int_0^1\Delta x^iP_{i,s}\,ds
		\right]_{t_0}^{t_1}
		-
		\int_{t_0}^{t_1}\int_0^1
		\bigl(P_{i,s}\Delta v^i+F_{i,s}\Delta x^i\bigr)\,ds\,dt.
	\end{aligned}
\end{equation}

Since we want an action-like expression, an integral of some function in time, we can define
\begin{equation}\label{eq:properFluxLagrangian}
	L_{\mathcal{F}}(x_a,v_a,x_b,v_b,t)
	:=
	\int_0^1
	\left[
	P_i(x_s,v_s,t)\Delta v^i
	+F_i(x_s,v_s,t)\Delta x^i
	\right]ds
\end{equation}
and obtain
\begin{equation}\label{eq:fluxLagrangianFlux}
	\mathcal{F}[x_a(t),x_b(t)]
	:=
	\int_{t_0}^{t_1}L_{\mathcal{F}}(x_a(t),d_t x_a(t),x_b(t),d_t x_b(t),t)\,dt.
\end{equation}
When the two paths have the same position endpoints, the boundary term vanishes and $\mathcal{F}[x_a(t),x_b(t)]=-\phi[\Sigma_{ab}]$.

In analogy with Lagrangian mechanics, we call $L_{\mathcal{F}}$ a \emph{flux Lagrangian} and its integral $\mathcal{F}$ the \emph{flux}. The former, unlike an action Lagrangian, compares two states at the same time; the latter compares two paths and, when the position endpoints coincide, equals minus the flux of configurations across the chosen surface between them. Moreover, the flux has a precise physical characterization in terms of configurations, which makes the name ``flux'' descriptive of both the geometry and the physics.

We now want to characterize the flux Lagrangian by two key properties. Note that exchanging the two states reverses $\Delta x$ and $\Delta v$ and replaces $s$ with $1-s$. Thus,
\begin{equation}\label{eq:fluxLagrangianAntisymmetry}
	L_{\mathcal{F}}(x_b,v_b,x_a,v_a,t)
	=
	-L_{\mathcal{F}}(x_a,v_a,x_b,v_b,t).
\end{equation}
In particular, $L_{\mathcal{F}}(x,v,x,v,t)=0$. This antisymmetry is a general property of the flux Lagrangian, as it does not depend on the dynamics, conservative or not.

The flux Lagrangian has another interesting property. If we differentiate it with respect to the second velocity we have
\begin{equation}
	\begin{aligned}
		\frac{\partial L_{\mathcal{F}}}{\partial v_b^j}
		&=
		\int_0^1
		\left[
		s\frac{\partial P_i}{\partial v^j}(x_s,v_s,t)
		\Delta v^i
		+P_j(x_s,v_s,t)
		+s\frac{\partial F_i}{\partial v^j}(x_s,v_s,t)
		\Delta x^i
		\right]ds.
	\end{aligned}
\end{equation}
When the states are the same, $\Delta x=\Delta v=0$. Therefore
\begin{equation}
	\left.
	\frac{\partial L_{\mathcal{F}}}{\partial v_b^j}
	\right|_{x_a=x_b=x,\;v_a=v_b=v}
	=
	P_j(x,v,t).
\end{equation}
By \ref{KE}, the map $v \mapsto P(x,v,t)$ is a global diffeomorphism between velocity and momentum at each fixed position and time. In analogy with Lagrangian mechanics, we call a flux Lagrangian with this property \emph{hyperregular}. Therefore, the flux Lagrangian constructed from the configuration flux is both antisymmetric and hyperregular.

By definition, then, since $\mathcal{F}[x_a(t),x_b(t)]=-\phi[\Sigma_{ab}]$ for paths with the same endpoints, the principle of stationary flux in phase space will lead to a stationary flux in kinematic form.

\subsection{From stationary flux to dynamics}

We now need to show that we can reverse the construction. First, we want to show that the principle of stationary flux leads to equations of motion for position and velocity. Suppose we identify states by position and velocity $(x^i,v^i)$ and that we are given a smooth antisymmetric hyperregular flux Lagrangian $L_{\mathcal{F}}(x_a,v_a,x_b,v_b,t)$. For convenience, define its diagonal derivatives by
\begin{equation}\label{eq:fluxLagrangianDiagonal}
	\begin{aligned}
		P_i(x,v,t)
		&:=
		\left.
		\frac{\partial L_{\mathcal{F}}}{\partial v_b^i}
		\right|_{x_a=x_b=x,\;v_a=v_b=v},\\
		F_i(x,v,t)
		&:=
		\left.
		\frac{\partial L_{\mathcal{F}}}{\partial x_b^i}
		\right|_{x_a=x_b=x,\;v_a=v_b=v}.
	\end{aligned}
\end{equation}

Consider a variation of the second path about the first,
\begin{equation}
	\begin{aligned}
	&x_a(t)=x(t),
	\qquad
	x_b(t)=x(t)+\epsilon\eta(t), \\
	&v_a(t)=d_tx(t),
	\qquad
	v_b(t)=d_tx(t)+\epsilon d_t\eta(t),
	\end{aligned}
\end{equation}
with $\eta(t_0)=\eta(t_1)=0$. By integrating by parts in the second term, we find
\begin{equation}
	\begin{aligned}
		\delta_b\mathcal{F}[x(t),x(t);\eta(t)]
		&:=
		\left.
		\frac{d}{d\epsilon}
		\mathcal{F}[x(t),x(t)+\epsilon\eta(t)]
		\right|_{\epsilon=0}\\
		&=
		\int_{t_0}^{t_1}
		\bigl(F_i\eta^i+P_i d_t\eta^i\bigr)\,dt
		=
		\int_{t_0}^{t_1}
		\bigl(F_i-d_tP_i\bigr)\eta^i\,dt.
	\end{aligned}
\end{equation}
Consequently,
\begin{equation}\label{eq:fluxLagrangianStationarity}
	\delta_b\mathcal{F}[x(t),x(t);\eta(t)]=0
	\text{ for all admissible }\eta(t)
	\iff
	d_tP_i=F_i,
	 v^i=d_tx^i.
\end{equation}
Expanding the total derivative of $P_i$, the equations become
\begin{equation}
	\frac{\partial P_i}{\partial v^j}d_tv^j
	=
	F_i
	-
	\frac{\partial P_i}{\partial x^j}v^j
	-
	\frac{\partial P_i}{\partial t}.
\end{equation}
By hyperregularity, $\frac{\partial P_i}{\partial v^j}$ is invertible and therefore the acceleration $d_t v$ is a function of $(x,v,t)$.

We find that the principle of stationary flux expressed through $\mathcal{F}$ for a hyperregular flux Lagrangian recovers the equations of motion. Note that stationarity here is taken in the second argument at coincident paths. It is not the variation of the identically zero functional $\mathcal{F}[x(t),x(t)]$ in both arguments together. We write $\delta_b \mathcal{F}[x(t),x(t)]$ to remind ourselves that the variation is taken on the second argument.

Lastly, we need to show that we can recover the dynamics in dynamical variables. Since $L_{\mathcal{F}}$ is hyperregular, we set
\begin{equation}
	q^i=x^i,
	\qquad
	p_i=P_i(x,v,t),
\end{equation}
and denote the inverse velocity map by $v^i=V^i(q,p,t)$. Define
\begin{equation}
	S^{q^i}(q,p,t)=V^i(q,p,t),
	\qquad
	S^{p_i}(q,p,t)=F_i(q,V(q,p,t),t),
	\qquad
	S^t=1.
\end{equation}
The stationary paths then satisfy $d_tq^i=S^{q^i}$ and $d_tp_i=S^{p_i}$. The constructed dynamical description has configuration counting form
\begin{equation}
	\omega=dq^i\wedge dp_i
\end{equation}
and configuration flux form
\begin{equation}
	w
	=
	dq^i\wedge dp_i
	-S^{p_i}dq^i\wedge dt
	+S^{q^i}dp_i\wedge dt.
\end{equation}

Thus a hyperregular flux Lagrangian supplies both the kinematic evolution and the correspondence between velocity and momentum. It also supplies a symplectic counting structure. This shows that \ref{CSD}+KE+TE-EOM is equivalent to \ref{CSK}+KE-STFL. It is remarkable that, apart from smoothness, hyperregularity and antisymmetry, no other requirement was put on the flux Lagrangian.

\subsection{Equivalent and proper flux Lagrangians}

Note that the variational equations depend only on the diagonal derivatives of $L_{\mathcal{F}}$. Two flux Lagrangians with the same diagonal derivatives, then, yield the same momentum map, counting structure and equations of motion. We therefore call two flux Lagrangians equivalent when they give the same functions $P_i$ and $F_i$ in Eq.~\eqref{eq:fluxLagrangianDiagonal}.

Given any hyperregular flux Lagrangian, we can extract $P_i$ and $F_i$ and use Eq.~\eqref{eq:properFluxLagrangian} to construct a new function $L_{\mathcal{F}}^{\mathrm{pr}}$. Its diagonal derivatives are
\begin{equation}
	\left.\frac{\partial L_{\mathcal{F}}^{\mathrm{pr}}}{\partial v_b^i}\right|_{\mathrm{diag}}
	=P_i,
	\qquad
	\left.\frac{\partial L_{\mathcal{F}}^{\mathrm{pr}}}{\partial x_b^i}\right|_{\mathrm{diag}}
	=F_i.
\end{equation}
Indeed, at coincident states the factors $\Delta x$ and $\Delta v$ vanish, leaving only the terms obtained by differentiating those factors. Thus $L_{\mathcal{F}}^{\mathrm{pr}}$ is equivalent to the original $L_{\mathcal{F}}$.

We call a flux Lagrangian \emph{proper} if it satisfies Eq.~\eqref{eq:properFluxLagrangian} with its own diagonal derivatives. Each equivalence class has exactly
one proper representative, since its defining expression is fixed by $P_i$ and $F_i$. The two constructions are therefore inverse in the following sense. Starting from dynamical variables, a counting form and an evolution satisfying KE, we obtain a proper flux Lagrangian that reconstructs the same data. Starting from an arbitrary hyperregular flux Lagrangian, we reconstruct those data and obtain its proper representative. This would be exactly the one recovered if we went to dynamical variables and then back to kinematic variables.

As with the action Lagrangian case, even in this general case the equations of motion do not uniquely identify a proper hyperregular flux Lagrangian. One can verify, for example, that for any smooth function $g(x,t)$
\begin{equation}
	P_i'=P_i+\partial_{x^i}g,
	\qquad
	F_i'=F_i+\partial_t\partial_{x^i}g
	+v^j\partial_{x^j}\partial_{x^i}g
\end{equation}
leads to
\begin{equation}	F_i'-d_tP_i'=F_i-d_tP_i,
\end{equation}
the same equations of motion. This is simply because conjugate momentum is defined up to a gauge. In a kinematic picture, this change of gauge will yield the same equations of motion.

\subsection{The conservative case}

We conclude by studying the conservative case and how the flux Lagrangian reduces to an action Lagrangian. Starting from dynamical variables, we already know that $w=-d\theta$, so the flux through the
surface between two paths with matching endpoints is minus the difference of their actions.
Under KE, these become actions expressed in kinematic variables. In fact, recalling that $L = p_i d_t q^i - H$,
\begin{equation}
	\begin{aligned}
		\mathcal{A}_H[\gamma] 
		&= \int_{t_0}^{t_1} ( p_i d_t q^i - H) dt  \stackrel{\text{KE}}{=} \int_{t_0}^{t_1} L(x^i(t), d_t x^i(t), t) dt = \mathcal{A}_L[x(t)].
	\end{aligned}
\end{equation}
The principle of stationary action, then, is exactly the principle of stationary action in phase space restricted by kinematic equivalence.

Conversely, starting with a hyperregular action Lagrangian $L$, we can define
\begin{equation}
	P_i(x,v,t)=\frac{\partial L}{\partial v^i},
	\qquad
	F_i(x,v,t)=\frac{\partial L}{\partial x^i}.
\end{equation}
The equations of motion become
\begin{equation}
	F_i-d_tP_i
	=
	\frac{\partial L}{\partial x^i}
	-
	d_t\frac{\partial L}{\partial v^i}
	=0,
	\qquad v^i=d_tx^i,
\end{equation}
which we recognize as the Euler-Lagrange equations.

By hyperregularity, $p_i=P_i(x,v,t)$ has a smooth global inverse
$v^i=V^i(q,p,t)$, with $q=x$. Define
\begin{equation}
	H(q,p,t)
	=
	p_iV^i(q,p,t)-L(q,V(q,p,t),t).
\end{equation}
Differentiating in dynamical coordinates, we obtain
\begin{equation}
	\begin{aligned}
		\frac{\partial H}{\partial p_i}
		&=
		V^i+
		\left(p_j-\frac{\partial L}{\partial v^j}\right)
		\frac{\partial V^j}{\partial p_i}
		=
		V^i,\\
		\frac{\partial H}{\partial q^i}
		&=
		\left(p_j-\frac{\partial L}{\partial v^j}\right)
		\frac{\partial V^j}{\partial q^i}
		-
		\frac{\partial L}{\partial x^i}
		=
		-F_i,
	\end{aligned}
\end{equation}
where derivatives of $L$ and the functions $F_i$ are evaluated at
$(x,v,t)=(q,V(q,p,t),t)$. Hence the reconstructed evolution satisfies
\begin{equation}
	S^{q^i}=\frac{\partial H}{\partial p_i},
	\qquad
	S^{p_i}=-\frac{\partial H}{\partial q^i}.
\end{equation}
Its configuration flux form is therefore
$w=-d(p_i\,dq^i-H\,dt)$ and is closed. This means that \ref{CSD}+KE+DI-SYMP is equivalent to \ref{CSK}+KEDI-STAC, proving the second equivalence of theorem \ref{thrm:bridge}. Since $DI-SYMP \implies TE-EOM$, the middle implication of theorem \ref{thrm:bridge} is also proved.

Let us now study the flux Lagrangian in the conservative case directly from its definition. We have
\begin{equation}
	S^{q^i}=\frac{\partial H}{\partial p_i},
	\qquad
	S^{p_i}=-\frac{\partial H}{\partial q^i}.
\end{equation}
Along the surface interpolation, these relations give
$v_s^i=\partial_{p_i}H$ and $F_{i,s}=-\partial_{q^i}H$.
Since $\partial_s t=0$, we have
\begin{equation}
	\begin{aligned}
		\partial_s H(x_s,P_s,t)
		&=
		\frac{\partial H}{\partial q^i}\Delta x^i
		+
		\frac{\partial H}{\partial p_i}\partial_sP_{i,s}\\
		&=
		-F_{i,s}\Delta x^i
		+v_s^i\partial_sP_{i,s}.
	\end{aligned}
\end{equation}
Substituting into the definition of the proper flux Lagrangian, and using integration by parts, yields
\begin{equation}
	\begin{aligned}
		L_{\mathcal{F}}^{\mathrm{pr}}(x_a,v_a,x_b,v_b,t)
		&=
		\int_0^1
		\left[
		P_{i,s}\Delta v^i+F_{i,s}\Delta x^i
		\right]ds\\
		&=
		\int_0^1
		\left[
		P_{i,s}\Delta v^i
		+v_s^i\partial_sP_{i,s}
		-\partial_s H(x_s,P_s,t)
		\right]ds\\
		&=
		\int_0^1
		\partial_s\left(P_{i,s}v_s^i-H(x_s,P_s,t)\right)ds\\
		&=
		\left[P_{i,s}v_s^i-H(x_s,P_s,t)\right]_{s=0}^{s=1}.
	\end{aligned}
\end{equation}
The boundary expression is precisely the action Lagrangian
\begin{equation}
	L(x,v,t)
	=
	P_i(x,v,t)v^i-H(x,P(x,v,t),t),
\end{equation}
viewed as a scalar function on extended phase space in kinematic coordinates. Since $(x_a,v_a)$ and $(x_b,v_b)$ are the endpoints of the interpolation in $s$, 
\begin{equation}\label{eq:fluxLagrangianDifference}
	L_{\mathcal{F}}^{\mathrm{pr}}(x_a,v_a,x_b,v_b,t)
	=
	L(x_b,v_b,t)-L(x_a,v_a,t).
\end{equation}
For a path and its variation, the flux is consequently
\begin{equation}
	\mathcal{F}^{\mathrm{pr}}[x(t),x(t)+\epsilon\eta(t)]
	=
	\mathcal{A}_L[x(t)+\epsilon\eta(t)]-\mathcal{A}_L[x(t)].
\end{equation}
This recovers stationary action as a special case. It remains to characterize conservative dynamics directly through the flux Lagrangian.

Note that Eq.~\eqref{eq:fluxLagrangianDifference} also gives an additional property. Writing $L_{\mathcal{F}}^{\mathrm{pr}}(\chi_a,\chi_b,t)$, where $\chi = (x, v)$ represents the pair of kinematic variables, we find
\begin{equation}\label{eq:fluxLagrangianAdditivity}
	L_{\mathcal{F}}^{\mathrm{pr}}(\chi_a,\chi_c,t)
	=
	L_{\mathcal{F}}^{\mathrm{pr}}(\chi_a,\chi_b,t)
	+
	L_{\mathcal{F}}^{\mathrm{pr}}(\chi_b,\chi_c,t).
\end{equation}
Equivalently, for every time interval $[t_0,t_1] \subset \mathbb{R}$, we have
\begin{equation}\label{eq:fluxActionAdditivity}
	\mathcal{F}^{\mathrm{pr}}[x_a(t),x_c(t)]
	=
	\mathcal{F}^{\mathrm{pr}}[x_a(t),x_b(t)]
	+
	\mathcal{F}^{\mathrm{pr}}[x_b(t),x_c(t)],
\end{equation}
for all paths, regardless of the endpoints. The above, in fact, means that 
\begin{equation}\int_{t_0}^{t_1}
\left[
L_{\mathcal{F}}^{\mathrm{pr}}(\chi_a(t),\chi_c(t),t)
-L_{\mathcal{F}}^{\mathrm{pr}}(\chi_a(t),\chi_b(t),t)
-L_{\mathcal{F}}^{\mathrm{pr}}(\chi_b(t),\chi_c(t),t)
\right]dt=0
\end{equation}
for all paths. This means the integrand must be zero at any given time. Intuitively, additivity means that the intermediate path cancels: the flux is determined by the initial and final paths. This mirrors the contour dependence of configuration flux in the conservative case.

With this property in mind, we can work the equivalence in reverse. Suppose we are given a proper hyperregular flux Lagrangian that satisfies additivity. Choose a reference kinematic state $\chi_r$ and
define
\begin{equation}
	L(x,v,t)
	:=
	L_{\mathcal{F}}^{\mathrm{pr}}(\chi_r,(x,v),t).
\end{equation}
Applying Eq.~\eqref{eq:fluxLagrangianAdditivity} to $\chi_r,\chi_a,\chi_b$ gives
\begin{equation}
	L_{\mathcal{F}}^{\mathrm{pr}}(\chi_a,\chi_b,t)=L(\chi_b,t)-L(\chi_a,t).
\end{equation}

If the above applies, the derivatives in the second set of state variables coincide with the derivatives of the action Lagrangian. Therefore, the momentum maps $P_i$ and the force maps $F_i$ are the same, regardless of whether they are defined on the flux Lagrangian or the action Lagrangian. This means that:
\begin{equation}
	\begin{aligned}
		dw=0
		&\iff
		L_{\mathcal{F}}^{\mathrm{pr}}(\chi_a,\chi_b,t)=L(\chi_b,t)-L(\chi_a,t)
		\text{ for some }L\\
		&\iff
		L_{\mathcal{F}}^{\mathrm{pr}}(\chi_a,\chi_c,t)
		=
		L_{\mathcal{F}}^{\mathrm{pr}}(\chi_a,\chi_b,t)
		+
		L_{\mathcal{F}}^{\mathrm{pr}}(\chi_b,\chi_c,t).
	\end{aligned}
\end{equation}
We find, then, that the conservative case is exactly when the proper flux Lagrangian is additive or, equivalently, is the difference of an action Lagrangian evaluated on two states. This means that KE-STFL $\impliedby$ KEDI-STFL $\iff$ KEDI-STAC, thus proving theorem \ref{thrm:kin}.

It should be stressed that additivity characterizes the conservative case only for proper flux Lagrangians. An equivalent flux Lagrangian need not itself satisfy this property, since its values away from coincident states are not fixed by the equations of motion and momentum map.

\section{Relationship to other variational principles}
\label{sec:othervarprinc}

The only reference we could find in the literature to a surface variational principle in the extended phase space for classical mechanics is due to Kochan~\cite{kochanstrings,kochan}. In that work, classical equations of motion are encoded in a dynamical two-form which, up to the sign convention, coincides with ours. No connection to the count or flux of configurations is established, and there is no expression of the principle in terms of a flux Lagrangian. Moreover, that result is not the main goal, but rather presented as a way to generalize the path integral of quantum mechanics to the nonconservative case. However, nonconservative dynamics in quantum mechanics does not conserve entropy, while the dynamics proposed in Refs.~\cite{kochanstrings,kochan} retains unitary evolution, thus conserving entropy. Such discrepancies are the reason we are looking for a fully physically motivated principle.

On the other hand, the only work we could find that introduces a two-path variational principle is due to Galley~\cite{galley} and developed further in Refs.~\cite{galleytsangstein,martindiego}. The principle is introduced as a modification of the principle of stationary action, which is taken as given, for general dynamics. While the formal part is very similar, the variational principle is described as acting on two histories, and it is presented as having a fundamental origin in quantum theory. Since we have shown that classical mechanics suffices to recover the principle, quantum theory cannot play a fundamental role in the origin of the principle. Moreover, Galley et al.~stress the decomposition of the flux Lagrangian into conservative and nonconservative parts, which, unfortunately, is not unique.

Our work, then, recovers these two approaches, reconciles them and puts them in a bigger, unified framework. That is, we recover Kochan's extended-phase-space formulation also in the case where KE fails, and specialize it to the kinematic formulation given by Galley et al. It also provides a better link to symplectic geometry and classical statistical mechanics, identifying the connection between the flux and the symplectic form, which is itself linked to the definition of entropy for statistical systems.

\section{Rayleigh dissipation}

We conclude by briefly calculating all the key objects for the case of a hyperregular action Lagrangian $L(x,v,t)$ and a Rayleigh dissipation function $R(x,v,t)$. The momentum and force maps are
\begin{equation}
	P_i=\frac{\partial L}{\partial v^i},
	\qquad
	F_i=\frac{\partial L}{\partial x^i}
	-\frac{\partial R}{\partial v^i}.
\end{equation}
Using the Hamiltonian associated with $L$, the equations of motion are
\begin{equation}
	\begin{aligned}
		\frac{dq^i}{dt}&=S^{q^i}=\frac{\partial H}{\partial p_i},\\
		\frac{dp_i}{dt}&=S^{p_i}=-\frac{\partial H}{\partial q^i}
		-\frac{\partial R}{\partial v^i},\\
		\frac{dt}{dt}&=S^t=1.
	\end{aligned}
\end{equation}
In dynamical variables, derivatives of $R$ are evaluated at
$(x,v)=(q,V(q,p,t))$. In the coordinate order $(q^i,p_i,t)$,
the configuration flux form has components
\begin{equation}
	w_{cd}=
	\begin{bmatrix}
		0 & \delta_i^j &
		\dfrac{\partial H}{\partial q^i}
		+\dfrac{\partial R}{\partial v^i}\\
		-\delta_j^i & 0 &
		\dfrac{\partial H}{\partial p_i}\\
		-\dfrac{\partial H}{\partial q^j}
		-\dfrac{\partial R}{\partial v^j} &
		-\dfrac{\partial H}{\partial p_j} & 0
	\end{bmatrix}.
\end{equation}

Using the interpolation $(x_s,v_s)$ defined above, the proper flux
Lagrangian becomes
\begin{equation}
	\begin{aligned}
		L_{\mathcal F}^{\mathrm{pr}}
		&=\int_0^1
		\left[
		\frac{\partial L}{\partial v^i}\Delta v^i
		+
		\left(
		\frac{\partial L}{\partial x^i}
		-\frac{\partial R}{\partial v^i}
		\right)\Delta x^i
		\right]_{(x_s,v_s,t)}ds\\
		&=L(x_b,v_b,t)-L(x_a,v_a,t)
		-\Delta x^i\int_0^1
		\frac{\partial R}{\partial v^i}(x_s,v_s,t)\,ds.
	\end{aligned}
\end{equation}
The first two terms are additive; the Rayleigh contribution
generally is not.

For quadratic dissipation with position-independent coefficients,
\begin{equation}
	R(v,t)=\frac12 C_{ij}(t)v^iv^j,
	\qquad C_{ij}=C_{ji},
\end{equation}
the integral gives
\begin{equation}
	L_{\mathcal F}^{\mathrm{pr}}
	=L(x_b,v_b,t)-L(x_a,v_a,t)
	-\frac12 C_{ij}(t)(x_b^i-x_a^i)(v_a^j+v_b^j).
\end{equation}

For example, a damped harmonic oscillator has
\begin{equation}
	L=\frac12 mv^2-\frac12 kx^2,
	\qquad R=\frac12 bv^2.
\end{equation}
Its equations of motion are
\begin{equation}
	\begin{aligned}
		\frac{dq}{dt}&=S^q=\frac{p}{m},\\
		\frac{dp}{dt}&=S^p=-kq-\frac{b}{m}p,\\
		\frac{dt}{dt}&=S^t=1.
	\end{aligned}
\end{equation}
Its configuration flux form and proper flux Lagrangian are
\begin{equation}
	w_{cd}=
	\begin{bmatrix}
		0 & 1 & kq+\dfrac{b}{m}p\\
		-1 & 0 & \dfrac{p}{m}\\
		-kq-\dfrac{b}{m}p & -\dfrac{p}{m} & 0
	\end{bmatrix},
\end{equation}
\begin{equation}
	L_{\mathcal F}^{\mathrm{pr}}
	=\frac m2(v_b^2-v_a^2)
	-\frac k2(x_b^2-x_a^2)
	-\frac b2(x_b-x_a)(v_a+v_b).
\end{equation}
Stationary flux gives $m\,d_t^2x+b\,d_t x+kx=0$.
For $b=0$, the flux Lagrangian reduces to the additive difference
of action Lagrangians.

\section{Summary}
\label{sec:summary}
We have derived from first principles a completely general, unified variational framework for both conservative and nonconservative dynamics of classical systems, defining a surface variational principle in classical phase space extended by time.  This framework can be used for systems described either by dynamical or kinematic variables.  Using reverse physics methodology, we have identified various relevant physical and mathematical conditions and established logical relationships among them, such that the physical commitments and assumptions entailed by different mathematical structures and expressions can be clearly understood.  We moreover believe that it is likely possible to remove the assumption of smoothness, with the result holding when the dynamics is simply once differentiable; however, we leave this to future work.

\bibliography{FluxPrinciple}

\end{document}